\documentclass[a4paper]{article}
\usepackage{amsmath}
\usepackage{amsfonts}
\usepackage{amssymb}
\usepackage{graphicx}

\usepackage{xcolor}
\usepackage{comment}

\usepackage[colorlinks = true,urlcolor  = blue,citecolor= blue]{hyperref}

\newenvironment{myitemize}
{ \begin{itemize}
    \setlength{\itemsep}{1pt}
    \setlength{\parskip}{0pt}
    \setlength{\parsep}{0pt}     }
{ \end{itemize}                  } 

\begin{document}
\pagenumbering{gobble}

\title{Code-Driven Law NO, Normware SI!}
\author{Giovanni Sileno\vspace{2pt}\\ 
   \normalsize \emph{Socially Intelligent Artificial Systems (SIAS) group}, \\ 
   \normalsize \emph{Informatics Institute}, \emph{University of Amsterdam} \\
   \normalsize \url{g.sileno@uva.nl} 
   }

\date{}

\maketitle

\begin{abstract}
With the digitalization of society, the interest, the debates and the research efforts concerning ``code'', ``law'', ``artificial intelligence'', and their various relationships, have been widely increasing. Yet, most arguments primarily focus on contemporary computational methods and artifacts (inferential models constructed via machine-learning methods, rule-based systems, smart contracts), rather than attempting to identify more fundamental mechanisms. Aiming to go beyond this conceptual limitation, this paper introduces and elaborates on ``normware'' as an explicit additional stance --- complementary to software and hardware --- for the \textit{interpretation} and the \textit{design} of artificial devices. By means of a few examples, I will argue that a normware-centred perspective provides a more adequate abstraction to study and design interactions between computational systems and human institutions, and may help with the design and development of technical interventions within wider socio-technical views.

\end{abstract}


\section{Introduction}


The concept of \emph{code-driven law}, i.e.~of ``legal norms or policies that have been
articulated in computer code'' by some actors with normative competence, has been convincingly elaborated by Hildebrandt \cite{Hildebrandt2020}. Its introduction has the merit to refocus the discussion on the role of artificial devices in the legal activity, rather than on ontological positions expressed under \emph{code-is-law} or \emph{law-is-code} banners, which are present, with various interpretations and changing fortunes, in the literature and practice of contemporary regulatory technologies, and technology-oriented legal scholarship (see the overview in \cite{Schrepel2022}).

According to Hildebrandt, code-driven law should be distinguished from
\emph{data-driven law}, i.e.~computational decision-making derived from statistical or other inductive methods, and from \emph{text-driven law}, i.e. the legal activity performed by humans by means of sources of norms such as statutory and case law.  A crucial difference between these forms of ``law'' is that the linguistic artifacts used in text-driven law are
characterized by open-textured concepts (e.g.~reasonableness, \textit{bona fide}, foreseeability, etc.) and multi-interpretability, and therefore carry inherently a potential of dispute that data-driven or code-driven law do not have (for statistical, or for logical closures). For this reason, code-driven
law could lead at best to \textit{computational legalism} \cite{Diver2021}, promoting \textit{legal certainty}, but not \textit{justice} (proportionality, distribution, etc.), nor
\textit{instrumentality} (i.e.~taking into account policy goals set by the legislature). In Hildebrandt's words: ``What code-driven law does
is to fold enactment, interpretation and application into one stroke,
collapsing the distance between legislator, executive and court.''

The present paper aims to respond to Hildebrandt's analysis, by
reframing some of the identified issues --- which are definitively observable in contemporary technologies --- in terms of technological gaps. \textit{What if we allow enactment, interpretation and application to be unfolded by design? }
{{Motivated by} this question we will form} a more nuanced view of the problem, highlighting spaces of computational research and practice that are (so far) for the most neglected, or partially addressed in different tracks, without enjoying the benefits of a unifying vision. In particular, the paper presents several arguments supporting the introduction of ``normware'' as an explicit third level --- besides hardware and software --- for the conceptualization and the design of artificial devices.\footnote{{The term ``normware'' was initially inspired by works on computational normative systems, traditionally focusing on legal knowledge artifacts. With this meaning, it was introduced in the workshop ``Normware: Modeling Norms and Automated Norm Application'', held at the Lorentz Center in Leiden (2016). Yet, we soon acknowledged that the traditional symbolic approach was insufficient to deal with structural problems raised e.g. by explainable AI and trustworthy AI \cite{Sileno2018}, and more in general with efforts concerning responsible computing, as for instance those addressed by the Responsible Internet proposal \cite{Sileno2021}. This paper widely extends and complements these works, constructing an overarching theoretical framework which was previously only sketched. 
}} 

The contribution of this work, although concerning a technical domain, is primarily conceptual. Its intended audience are both computer scientists working at the interface between computational and human systems, as well as legal scholars working on technological dimensions of society.  
Readers may find connections, echoes or at least analogies of parts of this exposition to seminal conceptual frameworks of authors as e.g. Deleuze's \textit{assemblage}  \cite{Deleuze1980}, Luhmann's theory of social systems \cite{Luhmann1995}, Latour's Actor-Network theory \cite{Latour2007}. This is not surprising: although our main goal here is identifying and mapping engineering dimensions relevant to the regulation of computational systems, the overall effort can also be seen as formulating aspects of social systems, albeit computational. Yet, the perspective and the assumptions underpinning the present work are rather different from those taken by these authors. We start here from engineering concerns, and therefore ontological considerations --- figuring an infrastructure, i.e. reusable computational components and their mechanisms of interactions --- are much stronger than epistemological ones. In purpose, this paper should rather be connected to contributions as Newell's \textit{knowledge level} \cite{Newell1982} (1982), which, reorganizing in an accessible conceptual framework various computational approaches based on reified explicit knowledge, paved the way to various achievements of symbolic AI, including the semantic web technology. As we will elaborate in the paper, the normware level encompasses the knowledge level, because it does not presuppose a symbolic reification, nor an intentional unity: it is meant to capture the more diverse forms of normativity expressed through computational systems, either reified as artifacts, or realized as processes.



\paragraph{{Outline of the paper}}

{The paper proceeds as follows. Section 2 starts from revisiting Hildebrandt's distinction \cite{Hildebrandt2020} between text-driven, data-driven, and code-driven law. The differences between the three, in particular in terms of contestability, may be less pronounced than what previously thought. From an engineering perspective, this observation opens up to reframing the problem differently, and we do so by introducing the notion of normware.
Section 3 will elaborate on normware as an artifact, taking both internal and external views, and identifying its three main functions: regulating behaviours, qualifications, and expectations. Section 4 will focus instead on normware as a process, filling an existing gap between the traditional ``control'' stance of hardware and software engineering to the ``guidance'' stance of policy- and norm-making. Section 5 will reply to possible objections derived from reductionist stances (``\textit{normware is just software}''), offering examples of application of normware as interpretative stance. Section 6 delineates research tracks whose relevance is further motivated by an explicit distinction (``\textit{normware is not software}''). }

\section{Types of law: a more nuanced distinction}


\subsection{Text-guided law?}
In recent work \cite{Hildebrandt2020}, Hildebrandt refers to ``text-driven law'' as the legal activity performed by humans by means of sources of norms expressed in verbal forms. Such an activity could not have been performed without a whole---traditionally non-computational---technological infrastructure related to textual artifacts, used for printing, distributing, archiving, methods for indexing and retrieving \cite{hildebrandt2015smart}. 
Yet, at a more fundamental level, text-based law inherently relies on language, and natural language functions differently than formal languages used in computational settings. Natural language is open to polysemy, it strongly relies on analogical constructs, and its use (in production and in interpretation) heavily depends on context. In some occasion, text in laws is intentionally left ambiguous, for instance when legislators, for reasons of political opportunity, prefer not to take a clear commitment towards a certain outcome. In other cases, multi-interpretability is a consequence of generalization. Law is expected to work in a \emph{differed} fashion, it does not apply only on known, expected cases, but it aims to be relevant also for cases that have not yet appeared. Humans in the loop are therefore inevitable, in particular with cases exhibiting new features or when changes are made to the institutional structure. Multi-interpretability also entails that different humans may frame the same case selecting and interpreting texts in different ways, opening to disputes and also to uncertainty with respect to the outcome of the decision. Part of the procedural mechanisms in court as well as of the jurisprudential debate have a crucial role in promoting legal certainty, but what matters here to us is that the relation between text and legal activity is not a direct one, and, insofar humans are involved, we should rather refer to ``\emph{text-guided law}''.\footnote{ 
Following a similar idea, if data-driven or code-driven inferential mechanisms are part of a support system for providing information or recommendations to human decision-makers on legal matters, it would be more proper to talk of ``\emph{data-guided law}'' or ``\emph{code-guided law}''.}

\subsection{Uncertainties in data-driven law}

Regardless of how data-driven inferential mechanisms are incorporated in the decision-making process, empirical alignment resulting from machine learning is generally performed against a certain training data-set, in principle collected and engineered to provide a sound \emph{statistical closure} to the target domain of application. The effectiveness of the trained model to achieve the desired outcome depends not only on the dataset, but also on the specific training method. Let us suppose, without loss of generality, that the data-set consists of a set of instances, each one associated with a label, as in a traditional supervised classification task. From a technical point of view, it is well known that good practices in data science and machine learning associate with critical questions, amongst which: \emph{Are the instances in the training data set representative of the domain upon which the model is applied?} \emph{Is the rationale behind the labelling still the same at runtime?} \emph{Is the chosen training method providing adequate, reproducible, robust results?} The third question can be directly associated to technical aspects, but the first two are prototypical examples of context-centred (and not algorithm-centred) inquiries, in particular concerning the assumptions of statistical closure.\footnote{See e.g. the RETRO-VIZ tool \cite{Debie2021}, estimating and explaining the trustworthiness of predictions on the basis of the statistical grounds on which the model has been constructed.} Thus, data-driven law can be contested, and indeed best practices entail that it should be scrutinized, even before setting additional requirements as e.g. explainability or transparency of the inferential process.

\subsection{Uncertainties in code-driven law}
The third form of legal activity identified by Hildebrandt is ``code-driven law'', for which provisions from various legal sources (e.g. legal norms, regulations, policies, agreements, contracts) have been operationalized via computational means. From a historical point of view, the idea of using machines with normative provisions traditionally dates back to Leibniz (\emph{Calculemus!}), and took more concrete forms with the advent of legal expert systems and various attempts at formalisation of law (e.g. \cite{McCarty1976,Sergot1986}), in part renewed during the semantic web technology heyday, and, in narrower forms, in contemporary RegTech (e.g. smart contracts), and cyber infrastructure research (e.g. data governance, consent management systems). Most distinctive assumptions of these contributions can be categorized within the domain of \emph{legal isomorphism} \cite{bench1992isomorphism}, susceptible to concerns on whether, to what extent{,} and how legal sources can be represented in computational artifacts, and normative reasoning automated by computational processes. At a strictly theoretical level, neglecting interpretative and tractability challenges, the problem becomes a purely logical one. Given a formal language, and the associated derivation mechanisms, soundness and completeness can be formally proven to verify a \emph{logical closure} guaranteeing valid inferences. Let us then suppose that our system is provided with \emph{structural knowledge} relative to norms in the form of \emph{rules}, and \emph{contingent knowledge} relative to the case in the form of \emph{facts}; within a formal system we can be certain that we utilize these resources to derive \textit{logically valid} conclusions. But are they correct in a practical sense? Theory and practice of computational legal theory acknowledge several critical questions in this respect: \emph{Were all the relevant fact and rules considered?}
\emph{Is the formalization correct?}
\emph{Is the reasoning tool robust?} As in the previous section, only the last question concerns proper technical dimensions (e.g. tractability, verification, security); the other two concern the selection of relevant contextual elements, and the choice of mapping from natural to formal language.

\subsection{Setting an alternative technological agenda}

Taking into account the critical questions identified in the respective fields of practice, we can conclude that neither data-driven nor code-driven law guarantee legal certainty. The premises on which the underlying computational devices are constructed and deployed will always be ``weak'' to some extent. 
Besides, software bugs, limitations or hacks {(malicious attacks)} may occur in  computational systems supporting both data-driven and code-driven law. Said differently, the idea of \textit{computational legalism}\footnote{{We refer to Diver \cite{Diver2021} for a detailed inquiry on the concept of \textit{computational legalism}. Related concepts can also be found in early works in AI \& Law, cf. \textit{mechanical jurisprudence} in \cite{Gardner1987}, or the ``legalistic child'' example discussed in \cite{Rissland1989}.}} works only in so far we forget all what may go wrong. The only way by which these computational devices may achieve legal certainty is by removing the possibility of contestation, which, although at times this is presented as a feature, it clearly brings risks of societally detrimental effects.

\paragraph{Naive computational justice, and its limits}
Perhaps, a more sustainable path is to reproduce in the computational realm procedures analogous to the ones introduced in human societies, which have proven value to deal with the legal (un)certainty of text-guided law. For instance, in order to satisfy principles of justice, such as \emph{proportionality} and \emph{distribution}, we may formalize those as meta-rules. In order to keep instrumentality with respect to contextual policy goals, we may require to make these goals explicit, and have them in the loop of the automated reasoning process. The resulting (naive) \textit{computational justice} stance is based on the belief that a perfect conceptualization will solve all the observed issues. Unfortunately, there are two serious hindrances for such an approach. First, we cannot be sure of any formalization, ever. The strongest support for this claim is not technical, but comes from the human dimension of the problem, the ``referent'' {(the entity referred to)} that needs to be formalized. For instance, the debates on ethical frameworks and human rights make clear that there is no such a referent, or, if it exists, it does so only on some contextual basis.\footnote{See e.g. Schlag's metaphor of law as ``hidden ball'' \cite{Schlag1996}.} Second, we may never be ready to have an explicit and complete articulation of policy-makers' purposes. Even imaging that this was possible, by means of perfect introspection and no errors in expression, making clear in detail all what some authority aims to achieve {(or does not aim to)} 
would entail to be transparent also on highly contestable issues, which in many occasions would be politically detrimental, and thus not to be expected.

\paragraph{Expecting wrongs, errors, redirections}
When both computational legalism and (naive) computational justice have been defeated, it seems there is no much hope for an advance in legal technologies. Fortunately, this is not true; new and sounder objectives can be set by starting from opposite assumptions. With respect to computational legalism, we should not suppose that things are correct by default (or eventually will be), but instead accept that things may always go wrong, at all levels. Under this new expectation, we may still accept to use legal technologies, as long as we also accept that \emph{people or other actors ought to be able to \textbf{appeal} algorithmic decisions}. Similarly, with respect to computational justice, we should accept that there may be always relevant directives not adequately articulated in the present computational artifacts. This entails that we need systems to integrate systematically feedbacks from higher-order courts, from jurisprudence or other competent normative sources. In order words, \emph{courts or other actors ought to be able to \textbf{quash} or \textbf{overrule} algorithmic decision-making processes}. 
Accepting these two principles leads us to technological innovations which have been neglected so far, aiming to address primarily computational instruments that operationalize appeal and quashing/overruling within the computational realm. Interestingly, such an infrastructure would support the systematical introduction of artificial agents, additionally to humans, to perform critical tasks like testing and auditing. 

{In short,} the core challenge of regulatory technologies is not captured only by  requirements as those expressed in software engineering terms, but by the construction of proper, functioning feedback loops, in the form of \emph{intervention points} for social actors provided with relevant competence. This level of abstraction makes clear that we are in fact in front of a problem of \emph{governance} (who has the power to do what, under which circumstances, and for what purpose), a problem which, when computational systems are involved as mediators, has also a technical dimension to be dealt with. This is where the concept of \emph{normware} fits into.


\section{Normware as artifacts}

\subsection{Directives concerning regulation}

In the most intuitive interpretation, ``normware'' refers to \textit{computational artifacts specifying norms}, i.e. intelligible directives concerning regulation of behaviour.
Common computational artifacts going under this definition would be for instance access control and usage control policies, machine-readable data-sharing agreements, digital consents, policies for digital infrastructure regulating e.g. routing, or data-sharing across domains, as well as behavioural policies for computational agents. Complementary to these more operational contributions, there is a large body of literature on \emph{normative systems}, consisting of a vast number of contributions on deontic logic, i.e. the logic of obligation, prohibition, and permission (see the overviews given in \cite{Aqvist2002,Hilpinen&McNamara2013}
), and fewer on alternative conceptual frameworks (e.g. Hohfeld's primitive normative concepts \cite{Lindahl1977,Jones1996,sileno2014b,Dong&Roy2017,Markovich2020,Pascucci2021}). 
Yet, even without referring to explicit normative concepts, or used explicitly for normative purposes, \textit{any} computer program has inherently a ``normative'' interpretation.

\paragraph{Imperative programs}

The most traditional form of programming artifacts is that of \emph{imperative programs}. Imperative programs consist of sequences of instructions, i.e.~\emph{actions that ought to be performed by the machine}.  Even \textit{machine code}, used to interact with hardware, is generally expressed in imperative terms. However, if the engineering objective for normware artifacts is to achieve some form of \emph{legal isomorphism} \cite{bench1992isomorphism} (or, more reasonably, a weaker version of it, i.e. targeting not the text of normative source, but a certain interpretation of the text, expressed by some expert), imperative programs cannot specify the diversity of institutional constructs that can be found in normative sources. For instance, concepts as prohibitions cannot be directly operationalized one-on-one to imperative programs, as we can define only what the machine has to perform, positively.\footnote{It is worth to note that Solidity, the most used language today to specify smart contracts, is an imperative programming language. This entails that, from a designer perspective, the main value of contemporary smart contract technology lies in providing by design relevant non-functional requirements (e.g. immutability) via distributed ledgers and decentralized execution, rather than a sound representational model to specify contractual mechanisms.} 

\paragraph{Declarative programs}

Several authors in the literature (see e.g. \cite{Kowalski2018,Governatori2018}) have observed that declarative programs may be a more suitable tool for representing norms, as norms may be specified in terms of \emph{constraints} \emph{that ought to be satisfied} (sometimes called ``goals''). Dedicated solvers can then be used to automate the process of finding solutions (e.g. behaviours) that satisfy these constraints. In most cases, goals can also be defined negatively, with the meaning ``\texttt{this} should not occur'', where \texttt{this}---simplifying---is some abstract or concrete situation (e.g. object or actor being in a certain state, an actor performing a certain action). 
More complex notations take also into account \emph{preferences} (also named \textit{weak goals}, or \textit{soft constraints}), i.e. the fact that goals may have a relative (and conditional) priority. Several computational methods have then been introduced to automatically derive the most preferred goals, under certain assumptions. A declarative approach is particularly relevant in domains in which goals can be conflicting, as priorititization enables or at least facilitates conflict resolution.

\subsection{Directives concerning qualification}

As soon as we climb the ladder of abstraction we are however confronted with the classic problems of semantics: \emph{What do we mean by} \texttt{this}\emph{?} (for language-mediated situations), correlative to \emph{What is} \texttt{this}\emph{?} (for concrete, direct situations). Part of these wide issues in philosophy and in symbolic AI goes under the name of \emph{qualification} problem \cite{Reiter2001}, concerning more generally the adequate expression of the preconditions required for an action (here, a certain naming) to be applicable. Interestingly, law attempts to partially solve similar challenges by means of \textit{constitutive rules} defining within the normative sources what \emph{count-as} what, or, complementarily, via procedural heuristics like evidential rules emerging from case law. 

Fortunately, being in the computational domain provides us with a privileged (but also illusive) starting point to tackle this problem, in comparison to a human-centred social domain. This is because hardware manufacturers physically hard-code the (local) semantics of primitive operations, consisting of logical and arithmetic operations, and specific instructions controlling memory or some input/output modules. As long as all possible ``\texttt{this}'' for our domain of concerns can be referred back to these physical primitives, or their logical counterparts, the qualification problem can be solved. 
However, because in any practical application we need to go beyond the computational boundary (there are always elements referring to or depending on what is outside the computational system), we cannot truly bypass the qualification problem, and this entails 
that we need terminological specifications provided by humans  to scope more adequately what is being regulated. These artifacts are also part of normware and aim to capture correct terminology, rather than correct (or wrong) behaviour.

\subsection{Directives concerning expectations}

Terminology is however not sufficient in itself. As a discipline, \textit{knowledge engineering} emerged as targeting more widely \emph{ontologies}: shared conceptualizations of a target domain expressed in a machine-processable form \cite{Gruber1993}. These conceptualization artifacts contain terminological knowledge, but they also include relationships attributed (in most cases by humans) to the target domain: e.g. taxonomies, partonomies, causal dependencies and logical constraints. Independently from the specific representational model associated to the various solutions introduced in the literature and by industry, all these artifacts aim to reify \emph{expectations} in symbolic forms. This was deemed relevant for computational regulative purposes, as most normative reasoning tasks require some form of \textit{world knowledge} \cite{Breuker2004}, e.g. to elaborate on causal chainings for responsibility attribution, to compare the impact of action vs omission of action, to evaluate foreseeability of events. {Besides, this elaboration concerns not only physical mechanisms/outcomes but also symbolic ones, as for instance with \emph{institutional powers}, specifying actions whose performance carries institutional meaning.} Unfortunately, it is also well-known that computational ontologies and more generally purely symbolic methods suffer from serious limitations (e.g. the infamous knowledge-acquisition bottleneck, the difficulty of alignment across ontologies, the lack of \emph{symbol grounding} \cite{Harnad1990}, the ramification and qualification problems, and so on), therefore ontologies (or similar types of technologies) cannot be deemed to be a sufficient mean. 

Now, ontological artifacts are meant to capture what is to be expected, or is at least possible in the target domain. Indeed, beside \emph{normativity}, the ``norm'' in normware captures also a \emph{normality} dimension. Interestingly, statistical methods are particularly suited to identify what is ``normal'' in a certain sample space, and the same applies with statistical machine-learning: latent spaces captured in machine-learning models are non-intelligible expression of regularities induced from observations. For instance, a ``cat'' classifier embeds some model of a prototypical cat,\footnote{{To be more precise, machine-learning models are traditionally divided in \textit{discriminative} or \textit{generative}, depending on the information principle applied for selecting the output. Only generative models are seen as maintaining a prototype concept of cat, as they allow us to produce a ``cat'' output if required so. Yet, discriminative models can still be seen as reifying an \emph{evidential process} meant to reach a conclusion on whether a given input is a cat.}}  time-series predictors embed models of temporal patterns, word-embeddings reify expectations of proximity between ``concepts'', large-language models (LLMs) predict next words given an initial textual context. May these artifacts have also a role within a computational framework centred on normware? 

\subsection{Devices intended to regulate}

In general, designed artifacts can be interpreted by knowing the function for which they have been designed. A door regulates the entrance to a certain building or room. A semaphore signals to vehicles, bikes or pedestrians whether they are allowed or prohibited to cross. Suppose the door to be also electronically controlled. In both cases (semaphore, and controlled door), we do not have access to the inner decision-making mechanism that brought to the current output state, but, as far as we can see this output (e.g. the door being closed, the light being red), and we know to what is referring to (e.g. allowing a certain action or not), we are able to interpret it, adapting our behaviour accordingly. The case of the door goes even further, as the device intervenes in the physical environment disabling our action without requiring our understanding. These examples support the idea that black-boxes as machine-learning models are artifacts expressing some form of normality/normativity, and as such they can be thought as normware artifacts too. In other words, \emph{under the lens of normware, there is no difference between code-driven and data-driven law.}\footnote{The distinction is in terms of implementation. In code-driven law, the underlying directives supporting a certain conclusion are explicit and made intelligible by relying on symbols whose meaning was (supposedly) stipulated at a certain moment, and expressed in the form of program; in data-driven law, these directives are tacit, expressed in the form of e.g. parameters of a neural network.} Both forms of computation are meant to regulate \emph{behaviours}, \emph{qualifications}, and \emph{expectations}. For instance, a model trained via reinforcement learning will provide as output actions that have to be performed conditionally to the observed input, a classifier will provide labels to input objects, a temporal predictor projections in the future of given states, an autoencoder a latent model of the domain from which e.g. to identify abnormal instances, a large language model a text completing a given prompt, and so on.


\section{Normware as processes}

\subsection{Regulation as control}
Institutional systems since ancient times are based on some assumption of intentionality (e.g. the cases in \cite[p.~102]{Watson2008}). The behaviour of intentional agents, by definition, can be interpreted in terms of \emph{purpose}, motives, and motivations. In many cases, however, weaker interpretations of action may also apply, which do not require the existence of a mental state associated with the intent, but still maintain a purposive character. For instance, the notion of \emph{control system} has a primary role in engineering disciplines, and issues of regulation (e.g. homeostasis) are central to the study of biological systems. A crucial difference between control systems and agents is that the latter are deemed to be able to generate their own goals autonomously (albeit with various levels of autonomy), while for the former, goals have been hard-coded by design (top-down), or by selection through some adaptive, possibly evolutionary process (bottom-up). 
The goal of an entity is then, in an \emph{ecological} perspective, the function of the entity within the environmental niche in which it is embedded, and transcends the entity itself. This is clear for designed artifacts, but natural entities also fall under similar considerations without taking a creationist stance (cf. the concept of \emph{teleonomy} introduced by Monod \cite{Monod1970} 
as counterpart to \emph{teleology}). Whether artificial or natural, designed or emergent, what counts in control is the existence of some reference (the \emph{target} of control) indicating a ``point'' or ``region'' in the space of possible behaviours or states of the system, which the entity is set to either achieve or avoid (the \emph{direction} of control).\footnote{More complex systems may additionally exhibit the direction ``to avoid to control'' (i.e. maintaining \emph{no-direction}) with respect to the target, cf. the concept of \emph{strong permission} in the normative systems literature \cite{Makinson1999,Alchourron1984}.} 
By defining directives by means of this control signature (target and direction), any regulative mechanisms can be abstracted from its implementation. 

\subsection{Indeterminacy of references and directives}

The legal activity and jurisprudential debates make clear that, for socially relevant phenomena, references cannot be defined once and for all, nor expressed entirely with a single artifact, and not even taking into account all normative artifacts (unless we take a strong legal positivist position). A sounder stance is to see these situational ``points'' as mere virtual placeholders for setting up and discussing socio-technical arrangements, contextually and historically dependent \cite{Gordon2022}. The correct functioning of these arrangements is enabled only by some common understanding or behavioural coordination amongst participants, which they have agreed, or accepted, or at least not contrasted. This acknowledgement entails that we cannot rely on the assumption that there is only one correct way to model (symbolically or sub-symbolically) what a ``cat'' is, in all circumstances; that all what a contract means can be reified into a single artifact; that there exists an abstract \emph{fairness point}, that can be discovered once and for all by some mathematical formula, and so on.

Indeterminacy of references entails indeterminacy of directives, but indeterminacy of directives may also occur because of \emph{antinomies}, i.e. when two or more conflicting directives exist. Legal systems generally introduce formal criteria to identify the relative priority of norms (e.g. \emph{lex specialis}, \emph{superior}, and \emph{posterior}), but it is generally in the \emph{hard-cases} that one can see that formal criteria are accompanied by underlying/implicit preferential structures, eventually embodied by the judges (``voices of the law''), and sometimes expressed as \emph{values} (see e.g. \cite{Bench-Capon2003} in the case-based reasoning literature), that are meant to balance or complement formal mechanisms (or their absence) to supposedly obtain a more just result. 
At a fundamental level, the primary source of these indeterminacies are the different positions, perceptions, knowledge, expectations, motivations, and preferences of the various participants to the social gathering, and this diversity in perspectives brings naturally to different qualifications.\footnote{See for instance the \emph{argumentative theory of reasoning} \cite{mercier2011}, humans capture from the world what it is deemed relevant to advance their interests, they do not strive for truth in absolute sense. Distinct selections lead consequently to distinct qualifications.}

All these observations apply similarly to components of distributed computational systems. Computational modules with the same function may be given different inputs (e.g. two similar sensors placed in different positions), and/or defined by distinct processes (either designed or trained) striving for different goals or value structures, and for this reason may produce conflicting output. Moreover, humans (with all their heterogeneity) intervene in the computational loop, for instance interacting via some interface with the computational system, or possibly feeding it with additional data and/or modifying its code. Conflicting directives cannot but be deemed systematic.

\subsection{Coordination and conflict resolution}

Taken as an atomic component, a piece of normware functions as a \textit{coordination mechanism}, just like a semaphore. The concern of such coordination may be \textit{prescriptive}, with normware artifacts meant to regulate other modules in a certain way (w.r.t. behaviour, or terminology); or \textit{descriptive}, in the sense that a certain coordination function is observed or ascribed to a certain component (in the form of qualifications, or expectations). 
However, the introduction of the norwmare level of computation is purposely meant to go beyond the single individual artifact, and to acknowledge the  ``ecological'' dimension of computation, i.e. the presence and entrenchment of several concurrent---possibly conflicting---coordination mechanisms within the same system. The ecological dimension becomes also evident when there are frictions between descriptive and prescriptive perspectives upon the same entity.

\paragraph{Dealing with failures}
We can identify two general families of failures due to conflicts: 
(i) components may not behave as expected; (ii) components concurrently sets incompatible directives. Failures of the first type generally trigger some control mechanism that 
either intervenes on the system (triggering a \emph{repair} action) or on the environment (\emph{remedy} action); or, in some cases, by dropping the directive which failed.\footnote{cf. the \emph{monitoring} and \emph{diagnosis} problem types in the problem-solving literature \cite{breuker1994}.} From a system's point of view, these violations need to be dealt with in order to maintain the encoded \emph{coupling} between system and environment. 

Failures of the second type are purely internal to the system, and require the inhibition or the removal of a number of components to restore directivity. These conflicts concern the \emph{viability} of the system. A system which is unable to restore directivity carries the risks of becoming idle, which often does not coincide with being neutral towards the two directions, but \emph{de facto} taking one of the two by default. For instance, in a context in which both a duty of $A$ and a prohibition of $A$ are issued, not doing anything means fulfilling the second directive.

{Note also that it is not necessary for failures to occur to trigger a system's response. Expectations have a crucial role for \textit{anticipating} conflicts, promoting preventive or preparatory behaviour. The various \textit{risk strategies} can for instance be thought as meta-directives dedicated to anticipatory functions, capturing the entrenchment between expectations and references targeted by directives.}




\paragraph{Higher-order indetermination}

The maintenance of processes of determination of references, and of resolution of conflicts requires in itself some reference/directive. This requirement links to the \emph{procedural} dimension of law. 
In turn, conflict resolution directives can also be contested, and dynamically changed. In the human realm, any contract needs to comply with contract law, which defines some procedural, formal requirements that make a contract valid. Contract law in itself has been issued by some legislative body, whose members---supposing a democratic system---were elected, with voting mechanisms defined in turn by some previous legislative body. The example is relevant for two reasons. First, it shows that determination/resolution of conflicts can be proceduralized in different, modular forms within the same normative system: e.g. aggregation of preferences by majority, formal criteria, intervention of humans at certain decision points (e.g. contract parties, judges, legislators, voters). Second, it raises the question of how a normative system should originate in order to unfold. 


\subsection{Primary and secondary drivers}

It is not in our scope to participate in the debate on what originates legal systems (e.g. Kelsen's \emph{Grundnorm}, or Schmitt's \emph{constituent power}, see e.g. \cite{Loughlin2014}). Being in the computational realm provides us with a more controlled grounding than the real world, because some pre-existing regulatory infrastructure is necessarily given: at some micro-level, components will eventually behave as control systems, with no expectation of failure. These elementary capacities of artificial devices (physical, symbolic) are generally associated to powers granted to users/developers. From this insight, we can  formulate a distinction between:
\begin{itemize}
    \item \emph{code-driven} computational entities, whose control flow perfectly aligns with the instructions provided by another component or by a human with adequate powers (e.g. the developer); 
    \item \emph{code-guided} computational entities, whose control flow includes mechanisms mediating the input instructions with other directives, which in some circumstances may completely override the input.
\end{itemize} 
Correspondingly, in terms of normware components, one can distinguish between \emph{primary drivers} (e.g. the ones hard-coded by the manufacturer); from \emph{secondary drivers}, which contingently contribute to further specify and constrain the system behaviour, on top of primary drivers.\footnote{{In principle, within secondary drivers, one could also envision some measure of normative strength or normative force, proportional to the level of influence of the directive to the actual system's behaviour.}} Deciding where the boundary lies between primary and secondary drivers is a fundamental design choice, which is obscured by traditional stances on computational artefacts.

\section{{Normware in the wild}} 


When directives are implemented as primary drivers we are in the domain of code-driven and data-driven law. The concept of normware gains however most of its value once secondary drivers are introduced in the system. Differently from hardware and software, which are primarily defined in terms of control, we expect \emph{a piece of normware not to be used for control, but primarily for guidance of the computational system.} This is a view that we apply naturally when we think of directives expressed to human agents (e.g. parents to kids, legislators/policy-makers to social participants), but not to artificial devices. 

Yet, because normware abstracts the underlying computational technology, it can be used as a framework to evaluate any concept and architecture associated to computational systems: Turing machines, smart contracts, operating systems, the Internet, knowledge-based systems, neural networks 
and other machine learning methods (e.g. ensemble methods), as well as more complex architectures. We will now review a few of these examples.

\begin{figure}
\centering
\scalebox{.12}{\includegraphics{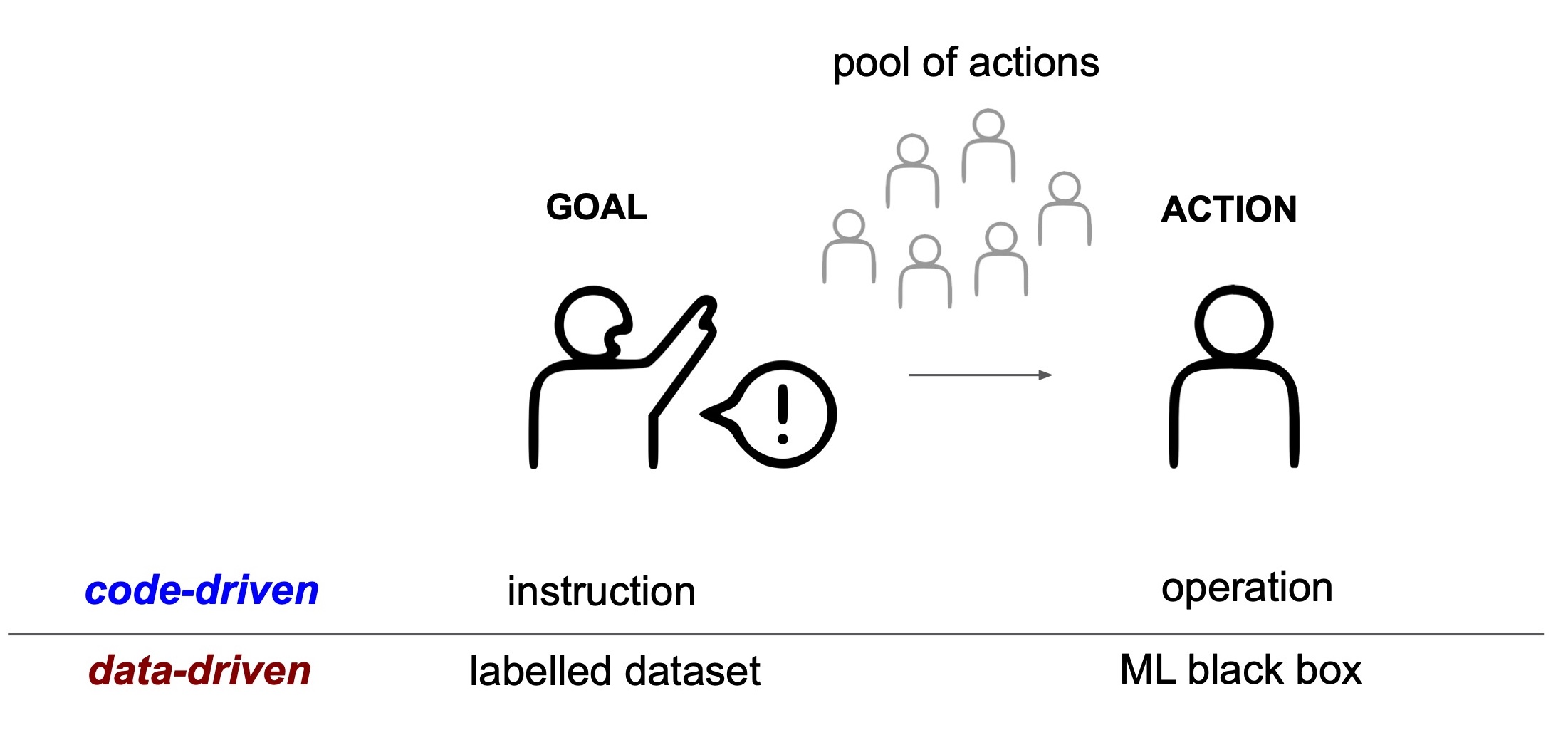}}
\caption{{Examples of first-order control: e.g. a user or developer provides a goal in the form of an instruction and the computer performs an action executing the associated operation. This association is hard-coded or logically derived in code-driven systems, or selected by some machine-learning method in data-driven systems.}}
\end{figure}

\subsection{{First-order control}}
{From theory, it is know that algorithms expressed in terms of \emph{control flow} of instructions---specifying the sequence of execution---can be reformulated in terms of \emph{control structure} of computational actors (but not vice-versa) \cite{Hewitt2013}. 
Consider for instance a ``printing agent'' implemented as a Turing machine.}\footnote{{A Turing machine is an essential model of computation consisting of: a \emph{tape}; a \emph{head}, which can read/write symbols on the tape, and move the tape left/right; a state \emph{register}; an \emph{action-table}. Seen as an artificial agent, the head of a Turing machine maps to the agent's physical body (sensors/actuators for reading/writing); the tape corresponds to the environment subjected to its physical actions; the state register corresponds to the mental state, whereas the action-table maps to automated behavioural responses.}}
{
Suppose this agent is designed to print on its tape, sequentially and indefinitely, the symbol it} {observes in its state register, overwriting what is recorded on the tape. Now imagine that, assuming there is no mechanical impediment, a second agent (possibly human) reads/write the state register of the ``printing agent''. The first machine, the controlled entity, can be seen as being \textit{subjected} to directives (about printing) issued by the second one, the controller entity. 
The same pattern can be observed between developers and computational systems in code-driven and data-driven systems (see Fig.~1): these systems cannot escape from what they have been instructed to do by the developers. Note how control structures like this one have a one-to-one mapping to \textit{power relationships}, and therefore can be captured by adequate normware specifications, although this effort adds impractical overhead with simple hierarchical constructs as the one expressed in this example. }


\begin{figure}
\centering
\scalebox{.12}{\includegraphics{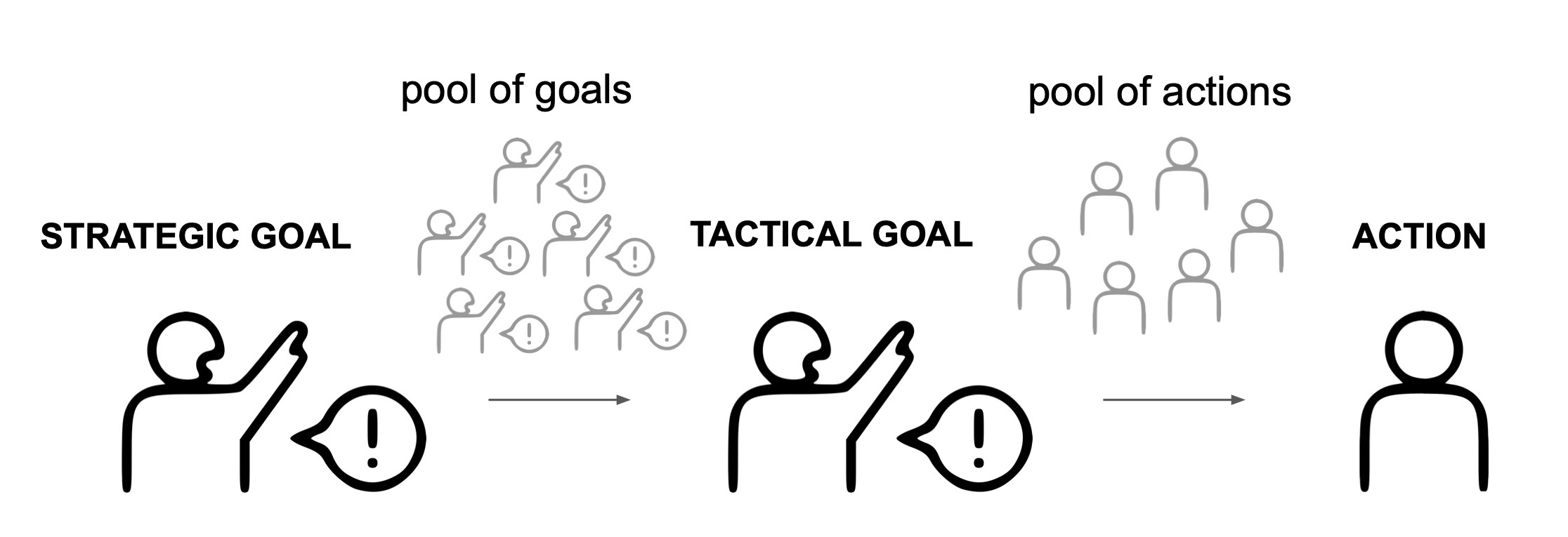}}
\caption{{Second-order control: the tactical goal is used to select an action amongst a pool of available actions, but the tactical goal in itself is selected from a pool of other goals depending on a strategic goal. The strategic goal captures a motivational context driving what the system is striving for.}}
\end{figure} 

\subsection{{Second-order control}}
{
Let us consider that the controller entity of the previous configuration is also a Turing machine. Its internal register may be modified in turn by a programmer, or by some other artificial entity enabled to so, and so on, realizing a hierarchy of control. While the controller/controlled construct is the prototypical domain of study of (first-order) \emph{cybernetics}, an additional control level brings us to the domain of second-order cybernetics. Similar three-levels views of systems are however introduced in other domains, as for instance: \textit{operations}, \textit{design}/\textit{development}, \textit{policy-making} activities in organizations \cite{Boer2013}, or \textit{operational}, \textit{tactical}, \textit{strategic} dimensions in decision-making (see Fig.~2).} 

{In previous work \cite{Sileno2018a}, we argued that contemporary challenges of \emph{explainable AI} can be framed as achieving reasonable abilities in justifying a certain decision, including being able to reject unacceptable, ``wrong" arguments. Similarly, \emph{trustworthy AI} can be associated to the requirement of not taking ``wrong” decisions, of performing “wrong” actions. In both cases, determination of unacceptability arises from conflicts between tactical and strategic levels, or between tactical and operational level.} 
{As an example, let us take the case of a computational model used for a classification task. Labels given with the training set are generally deemed to be consistent, but in reality this may be not the case. For instance, in administrative activities, the policies regulating the provided services may change, bringing to different categorization of similar cases. New behavioural trends may also emerge in the population, which were not captured in the original training dataset, thus causing inaccuracies in prediction. Most common machine-learning architectures lack a proper diagnostics feedback for these type of misalignments: their architecture is essentially one of first-order control. }
{However, it is known that second-order feedback mechanisms are necessary for satisfying requirements of organizational intelligence (e.g. Beer's \textit{viable system model} \cite{Beer1995}).} {This suggests that reappraising the control structure of machine learning components within a larger normware network can be a valuable exercise. 
Answering to the question \textit{what is controlling what} can help in identifying and then addressing deficiencies (for further examples see \cite{Sileno2018a}).}

\begin{figure}
\centering
\scalebox{.12}{\includegraphics{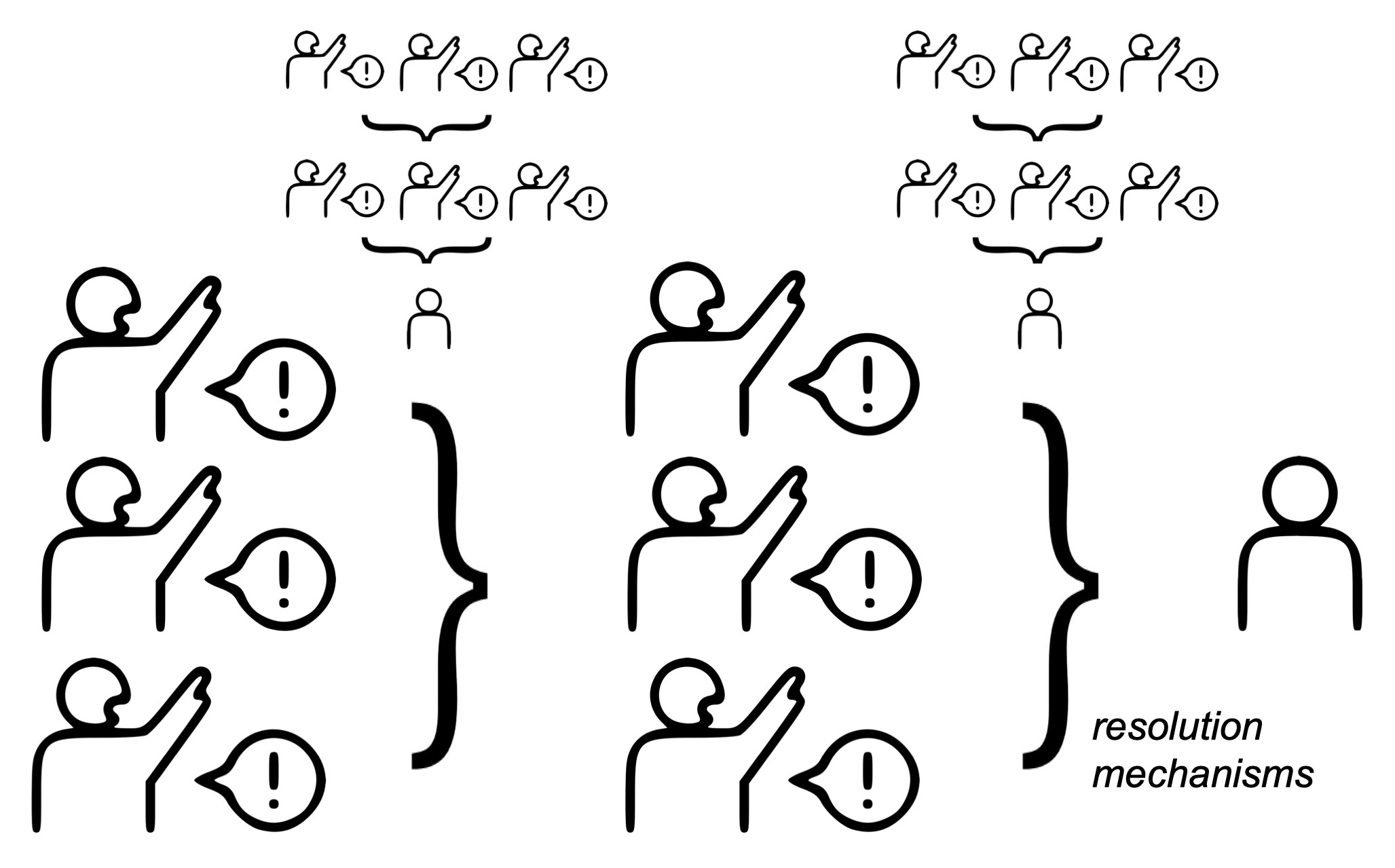}}
\caption{{Plural second-order control: competing directives at strategic and tactical levels are aggregated/simplified by resolution mechanisms, which in turn may be resulting from previous resolution actions.}}
\end{figure}

\subsubsection{Decentralizing control}{Traditional programs, as well as inferential models constructed via machine learning, are generally conceptualized to be executed as single, unitary computational processes. Similarly, the knowledge level presented by Newell assumes a \textit{single} agent whose ``law of behaviour'' follows a  ``principle of rationality''  \cite{Newell1982}. Or still, cybernetics. Introduced as the study of regulation mechanisms ``in the animal and the machine'' (Wiener, \cite{Wiener1948}), even when approached in second-order forms, and applied to interpret collective entities as organizations, cybernetics maintains a fundamentally \textit{organicist} view of systems.  This stance has obvious limitations, as it cannot support adequately the understanding of ecological systems (in our case, complex computational architectures).}

{In contrast, \textit{networking} is plausibly the computational domain that realizes at best Deleuze and Guattari's \textit{assemblage} view of systems \cite{Deleuze1980}. Even if most of the attention on ``responsible computing'' goes nowadays to machine-learning-based AI (and in particular generative AI, including chatbots based on LLMs), networking is also relevant to this challenge, because transporting data is not a neutral endeavour, for the consequences it may cause or facilitate at individual and collective levels. This is made clear for instance in the Responsible Internet proposal \cite{Hesselman2020}. As argued in a subsequent elaboration \cite{Sileno2021}, the distributed and decentralized nature of networks intrinsically requires instruments for}:
\begin{myitemize} 
\item dealing with a plurality of policies (of higher and lower abstractions, spatially and temporally located) set by users, network operators, and/or the various governance bodies;
\item dealing with a plurality of observations/expectations: by changing point of view and observational scale any target phenomenon typically maps to different interpretative and predictive models;
\item specifying mechanisms of creation, update, removal of policies and of expectations, assigned to the various stakeholders;
\item specifying their entrenchment in operationalization.
\end{myitemize}
{
The resulting view is a call for technologies enabling \textit{plural second-order control} (see Fig.~3), whose participatory structures may resemble, if not put in continuity, with those operating in human social systems.}

{Interestingly, the normware level provides a conceptual framework integrating these requirements in their entirety. The first two requirements concerns artifacts that are meant to regulate behaviours, qualifications, and expectations, addressed in section 3. The third requirement is about institutional powers assigned to parties, encoding the backbone control structure of the infrastructure. The fourth requirement consists of meta-directives specifying mechanisms of conflict resolution, including anticipation. These last two were addressed in section 4, and capture what is needed to develop a computational counterpart of \textit{governance} processes.}

\section{Research perspectives}
{Having set the conceptual horizon of normware, we will now briefly elaborate on two main research tracks relevant for a ``normware-centred'' engineering: how to specify normware (for individual artefacts), and how to operationalize normware (as an ecological process), focusing in particular on conflicts.}

\subsection{Specifying code \label{sec:which-code}}
Although the paper argues for going beyond a code-driven law perspective, code still plays a central role in code-guided law. To enable humans to specify normware in an intelligible and traceable way, we have plausibly to strive for some adequate mapping from (contextual interpretations of) sources of norms to some computational formal language. Since the first pioneering experiments with TAXMAN for US corporate tax law \cite{McCarty1976}, and the modelling of the British Citizenship Act with a logic program \cite{Sergot1986}, several solutions for specifying normative artifacts (norms, contracts, policies) in a computational processable way have been presented in the literature. 
Amongst the most recent efforts we acknowledge e.g.  LegalRuleML \cite{Palmirani2011,Lam2019}, PROLEG \cite{Satoh2011}, InstAL \cite{Padget2016}, ODRL \cite{Iannella2018,DeVos2019}, Symboleo \cite{Sharifi2020}, FLINT/eFLINT \cite{VanDoesburg2016,VanBinsbergen2020}, Logical English \cite{Kowalski2021}, Catala \cite{Merigoux2021}, Blawx \cite{Morris22}, Stipula \cite{crafa2022stipula}, and DPCL \cite{Sileno2022}. 
Legal core ontologies (e.g. LKIF-core \cite{Hoekstra2007}, UFO-L \cite{Griffo2018}) have also been proposed to systematize concepts and relationships relevant to normative reasoning. No solution amongst those has achieved general acceptance, even less in deployment. More unexpectedly, there exists no common ground (nor representational, nor computational) enabling a comparison between these solutions. Complementary to these efforts, there exist a number of industrial, domain-specific standards, which---in contrast to the above---are widely used, e.g. XACML (for access control) \cite{XACML3},  BGP policies (for routing in networking) \cite{Caesar2005}, Protune, Rei, Ponder, TrustX (for cloud services) \cite{Duma2007}. These solutions considers only few or even no explicit normative concepts; their concerns is not modelling what is in the law or in executive policies, but of having effective and computationally efficient instruments of regulation (for security, privacy, etc.). Similar considerations apply to artifacts to specify terminology and expectations like computational ontologies, logic programs, or functionally similar entities.

Law-, policy-, and more in general knowledge-centred software is evaluated against both a language dimension (determining what types of knowledge we can specify and how), and a computational dimension (determining what we can actually do with these knowledge artifacts). Approaches giving priority to logical aspects result in well-defined semantics, which enables the possibility of using formal verification to guarantee the correctedness of the system. However, despite efforts in such sense, there are strong reasons to believe we may never reach an agreed ``standard'' semantics. First, there exist an endless number of semantic dimensions that play a role in law (deontic, potestative, action/dynamics, priority/defaults, intentionality). Standards seem a feasible objective only in narrow sub-domains (e.g. primarily declarative, as for instance reasoning on the amount of due taxes), without capturing meta-norms and then overlooking at known properties of human norms like \textit{defeasibility}. Not surprisingly, for reasons as reliability, reusability, and better performance, industrial formalisms have much simpler representational models. In many cases, one needs not to introduce explicit normative categories \cite{BenchCapon1989,Logrippo2007,Kowalski2018}. 
Second, any semantics chosen can be seen as a commitment, and any commitment can be argued and contested under certain conditions (as we already observed in section 2.3). Only hardware manufacturers are exempt from these considerations, as they have hard-coded physically the local semantics of their components. 

Besides the formal verification aspect, a central (and overlooked) problem in making norms executable is their \textit{social operationalization}: i.e. of  transforming their declarative aspects into procedures. Not in the programming sense, but in the socio-legal sense: addressing governance mechanisms entails {designing the right \textit{intervention points} (who has the power to do what, and under which conditions}.
This aspect is also a matter of commitment. {For instance, a duty can be operationalized with different types of enforcement, each placing (or not placing) different burdens and incentives upon the social participants.}

Accepting that law-encoding software relies on choices of commitments---concerning both semantics and social operationalization---suggests also that these commitments should be (primarily) ascribed to the organizations using this software. The presence of a strong standard (e.g. issued by a central authority) may be used by parties to deny responsibility, and propagate systematic errors from central nodes. 
{In these conditions, the only viable collective goal is to strive for a core integration standard, i.e. a sort of \textit{knowledge interchange format}, an agreement on a minimal informational model encompassing architectural aspects as the functions described in section 3 and 4.
}



\subsection{Resolving conflicts}
Dealing with pluralism means not only enabling the existence of several concurrent, potentially competing components, but also to have instruments to resolve such conflicts. Because these instruments are also coordination artifacts, they belong to the normware level too. Indeed, one could encode in a normware artefact e.g. the formal meta-rules determining the priority between laws, or voting mechanisms, or other relevant institutional patterns.

Relevant conflict resolution mechanisms have been researched in various disciplines, providing in principle off-the-shelf methods that can be reused in a normware-based computational system. For instance, aggregation of preferential structures, e.g. by voting mechanisms, have been studied extensively in \emph{computational social choice} (see e.g. \cite{moulin2016}), as well as other branches of applied mathematics \cite{Bloch2021}. Directives can also be seen as public arguments. \emph{Formal argumentation theory} (e.g. \cite{baroni2018}) focuses on attacks (and/or support) relationships between individual arguments, and provides several methods to evaluate the strength of arguments based on the whole argumentation system. The various forms of \emph{defeasible logics} introduce rule priorities in the derivation process, approaching conflict resolution in hierarchical terms; this is particularly relevant in the legislative context, for instance to model the hierarchy of law, as well to capture priorities between values underpinning norms. Defeasible reasoning accounts of precedent are also studied in the \textit{case-based reasoning} research track \cite{Rissland2005,Horty2012}. However, the ecological perspective of normware opens also to a completely different way to look at the problem, via \emph{evolutionary computation} techniques: a normware ecology may be seen as a population subjected to artificial selection, with fitness defined dually to conflict. 

Which of these techniques will be actually useful in practical settings (and under which conditions) is clearly an open question, and out of the scope of this paper. 
Yet, it is relevant to observe we have found an unexpected ground of integration for independent efforts scattered across various computational sub-fields.


\section{Conclusion}

The concept of normware brings to the foreground that both institutional systems (particularly in their procedural aspects) and computational systems are in fact \emph{information processing systems}. Indeed, any institutional system transcends its individual social participants, and, to the extent that such a system is designed (i.e. its form has been deliberately chosen by rational agents), it can also be seen as an artificial one.

{Now, bureaucratic alienation in formal organizations (e.g. public administrations) is a well known phenomenon, both for participants in these organizations  
and for the various stakeholders interacting with them. Computational systems can be related to an analogous, plausibly even more intense alienation. Computers process symbols or numbers without any clue on their meaning further on in the decision process. 

From a systematic point of view, no ``intelligent'' system can be expected without: (i) designing adequate feedback loops (and the associated intervention points); (ii) taking adequately into account a plurality of drivers and of contexts; (iii) being able to identify and overcome the conflicts that will naturally occur. These three points summarize the most distinctive matters we have elaborated in this paper, investigating the various aspects of normware. Engineering perspectives at  an hardware (physical artefacts/processes), or software (symbolic artefacts/processes) levels do not provide the right abstraction to discuss about governance  (i.e. coordination artefacts/processes) within the computational realm. }

The continuity between institutional and computational systems however does not imply that accepting more sound computational forms of law will remove the need for human law. On the contrary, as human intervention is required to repair issues due to bureaucratic alienation, the best way to define their relationship (in the human and in the computational domain) is as the relationship that higher-courts have towards lower-courts, ensuring that humans will always remain in control and mechanisms like \emph{appeals} (against decisions) and \emph{quashing/overruling} (against decision-making processes) become systemic. 

In fact, a core conclusion of the present contribution is that \emph{the socio-technical challenge we are facing is not mechanizing law, but introducing legitimate normative processes within the computational realm}. Discussions on legality and legitimacy cannot solve the challenges on designing/developing technical possibilities supporting both; normware-centred discussions are instead required to advance the latter dimension. 

\bibliographystyle{unsrt}
\bibliography{references}




\end{document}